\documentclass[a4paper]{jpconf}
\usepackage{graphicx}
\begin{document}
\title{Remaining inconsistencies with solar neutrinos: can spin flavour 
precession provide a clue?}

\author{\underline{Jo\~{a}o Pulido}$^1$, C R Das$^1$,
Marco Picariello$^2$}

\address{$^1$ Centro de F\'{\i}sica Te\'{o}rica das Part\'{\i}culas, IST,
Av. Rovisco Pais 1049-001 Lisboa, Portugal 
}
\address{$^2$ INFN, Lecce and Dipartimento di Fisica, Universit\'{a} di
Lecce, Via Arnesano, I-73100, Lecce, Italia}

\ead{pulido@cftp.ist.utl.pt, crdas@cftp.ist.utl.pt, Marco.Picariello@le.infn.it}

\begin{abstract}
A few inconsistencies remain after it has been ascertained that LMA is the
dominant solution to the solar neutrino problem: why is the SuperKamiokande
spectrum flat and why is the Chlorine rate prediction over two standard
deviations above the data. There also remains the ananswered and important
question of whether the active neutrino flux is constant or time varying.
We propose a scenario involving spin flavour precession to sterile neutrinos
with three active flavours that predicts a flat SuperK spectrum and a
Chlorine rate prediction more consistent with data. We also argue
that running the Borexino experiment during the next few years may provide
a very important clue as to the possible variability of the solar neutrino flux.

\end{abstract}

Ever since the first KamLAND results were released \cite{Eguchi}, the large
mixing angle solution (LMA) \cite{Anselmann} has been generally accepted to be the 
dominant one for the solar neutrino problem. Nevertheless a few intriguing questions
remain open in this area. In fact it is customary to assume that the active solar 
neutrino flux is constant, although the possibility of its time variation is still
a challenging issue to be investigated \cite{Pandola}, \cite{Sturrock} both 
experimentally and theoretically. 
Moreover, the observed flatness of the SuperKamiokande spectrum \cite{Fukuda} and 
the Chlorine 
rate prediction lying at more than 2$\sigma$ above the data \cite{Cleveland}
are disturbing facts
providing evidence that LMA may not be the ultimate solution.

Based on these motivations, we developed a model which includes a light sterile 
neutrino in addition to the three active ones \cite{Das} which communicates with these 
by means of transition magnetic moments. Somewhere inside the sun, in a location
determined by the balance between the solar density and the mass square difference
between the sterile neutrino and the active ones, a resonant transition may occur
converting the latter into the sterile in the presence of a large magnetic field. 
Hence, besides the solar and 
atmospheric mass differences, a third one is introduced implying another
resonance in addition to the LMA one. For simplicity no extra vacuum mixing angle 
is assumed, so that the matter Hamiltonian which includes vacuum propagation is 
\cite{Das}
\begin{equation}
H_M=\left(\begin{array}{cccc} \frac{\Delta m^2_{01}}{2E} & \tilde \mu_1 B
& \tilde \mu_2 B & \tilde \mu_3 B\\
\tilde \mu_1 B & V_n+V_c u^2_{e_1} & V_c u_{e_1}u_{e_2} & V_c u_{e_1}u_{e_3} \\
\tilde \mu_2 B& V_c u_{e_1}u_{e_2} & \frac{\Delta m^2_{21}}{2E}+V_n+V_c u^2_{e_2}
& V_c u_{e_2}u_{e_3} \\ \tilde \mu_3 B& V_c u_{e_1}u_{e_3} & V_c u_{e_2}u_{e_3} &
\frac{\Delta m^2_{31}}{2E}+V_n+V_c u^2_{e_3} \end{array}\right).
\end{equation}
Here $V_c$, $V_n$ are the charged and neutral current interaction potentials, 
$\tilde \mu_{1,2,3}$ are transition magnetic moments between mass eigenstates 
0 and 1, 2, 3 and $u_{e_{i}}$ are the first row entries of the $(3{\rm x} 3)$ PMNS 
matrix. The newly introduced mass squared difference $\Delta m^2_{01}$ is 
smaller than $\Delta m^2_{21}$ and so the spin flavour precession resonance 
lies closer to the solar surface than the LMA one.

We have used the numerical integration of the Hamiltonian equation and 
considered the following two field profiles which are approximately complementary 
to each other \cite{Das}

\noindent {\it Profile 1}
$$ B_1=B_0/\cosh[6(x-0.71)]~~(0<x<0.71),~~~~
B_1=B_0/\cosh[15(x-0.71)]~~(0.71<x<1)$$
{\it Profile 2}
$$B_2=B_0/[1+\exp[10(2x-1)] \quad 0<x<1.$$

The experimental data on all rates (Cl, Ga, SK rate and spectrum,
SNO rates and spectrum, Borexino \cite{Bellini}) were used to assess the
quality of the fits with the following $\chi^2$ standard definition
$$\chi^2=\sum_{j_{1},j_{2}}({R}^{th}_{j_{1}}-{R_{j_{1}}}^{\exp})\left[{\sigma^2}
(tot)\right]^{-1}_{j_{1}j_{2}}({R}^{th}_{j_{2}}-{R_{j_{2}}}^{\exp}).$$
Using the experimental values of $\Delta m^2_{21},~\Delta m^2_{atm},~\theta_{12},~\theta_{atm}$ 
and $B_0$ (field at the peak), $\Delta m^2_{01}$ as free parameters, we obtained the fits 
for profile 1 shown in table I.
\begin{center}
\begin{tabular}{cc|cccccccccc} \\ \hline \hline
$B_0(kG)$ & $\sin \theta_{13}$ & Ga & Cl & SK & $\rm{SNO_{NC}}$ & $\rm{SNO_{CC}}$ & $\rm{SNO_{ES}}$ &
$\!\!\chi^2_{rates}\!\!$ & $\chi^2_{{SK}_{sp}}$ & $\chi^2_{SNO}$ & $\chi^2_{gl}$\\ \hline
 & 0  & 67.2 & 2.99 & 2.51 & 5.62 & 1.90 & 2.49 & 0.07 & 42.7 & 57.2 & 99.9 \\
0 & 0.1 & 66.0 & 2.94 & 2.49 & 5.62 & 1.87 & 2.46 & 0.30 & 42.1 & 55.2 & 97.6 \\
 & 0.13 & 65.0 & 2.90 & 2.46 & 5.62 & 1.84 & 2.44 & 0.62 & 41.7 & 53.7 & 96.0 \\
\hline
  & 0  & 66.4 & 2.82 & 2.32 & 5.37 & 1.76 & 2.31 & 0.20 & 37.6 & 46.0 & 83.8 \\
140  & 0.1  & 65.3 & 2.77 & 2.29 & 5.37 & 1.73 & 2.28 & 0.53 & 37.9 & 44.9 & 83.3 \\
  & 0.13  & 64.3 & 2.72 & 2.27 & 5.37 & 1.70 & 2.25 & 0.95 & 38.4 & 44.1 & 83.4 \\
\hline
\end{tabular}
\end{center}
{\it{Table I. $\chi^2$'s and rates (in SNU for Cl and Ga experiments, in $10^6 cm^{-2}s^{-1}$ for
SK and SNO).}}

\vspace{0.18cm}

For the best fit we found $B_0$=140 kG, $\Delta m^2_{01}=1.25\times 10^{-7}eV^2$ 
with transition moments between flavour eigenstates $\mu_{(\mu,\tau)s}=
1.4\times 10^{-12}\mu_B$ and $\mu_{es}\leq \mu_{(\mu,\tau)s}$.

The corresponding results for the SuperKamiokande and Borexino spectra in 
the case of profile 1 are shown in figs. 1 and 2. For profile 2 the quality 
of the fits and the results are much the same, except that the best fit is 
obtained for $B_0=0.75MG$, $\Delta m^2_{01}=2.7\times 10^{-6}eV^2$.

From table I and figs. 1, 2 a clear preference of the data for a sizable magnetic
field is apparent. In fact 

\begin{itemize}
\item the SuperKamiokande spectrum becomes flat
\item the rate prediction for the Chlorine experiment strongly improves (instead of
a 2$\sigma$ discrepancy, a prediction within 1$\sigma$ of the data is obtained).
\end{itemize}

Furthermore no conclusion can be drawn on the value of $\theta_{13}$.
As for the Ga rate, vanishing and sizable fields are equivalent, as both classes
of predictions lie within 1$\sigma$ of the central value. Regarding Borexino it is 
unclear whether it can favour either a negligible or a sizable solar magnetic 
field owing to the size of the experimental errors. Likewise no conclusion is 
obtained regarding the magnitude of $\theta_{13}$ in this case. So as this stage
Borexino results are consistent with both a vanishing and a large field.

\begin{figure}[htb]
\begin{minipage}{18pc}
\includegraphics[width=18pc]{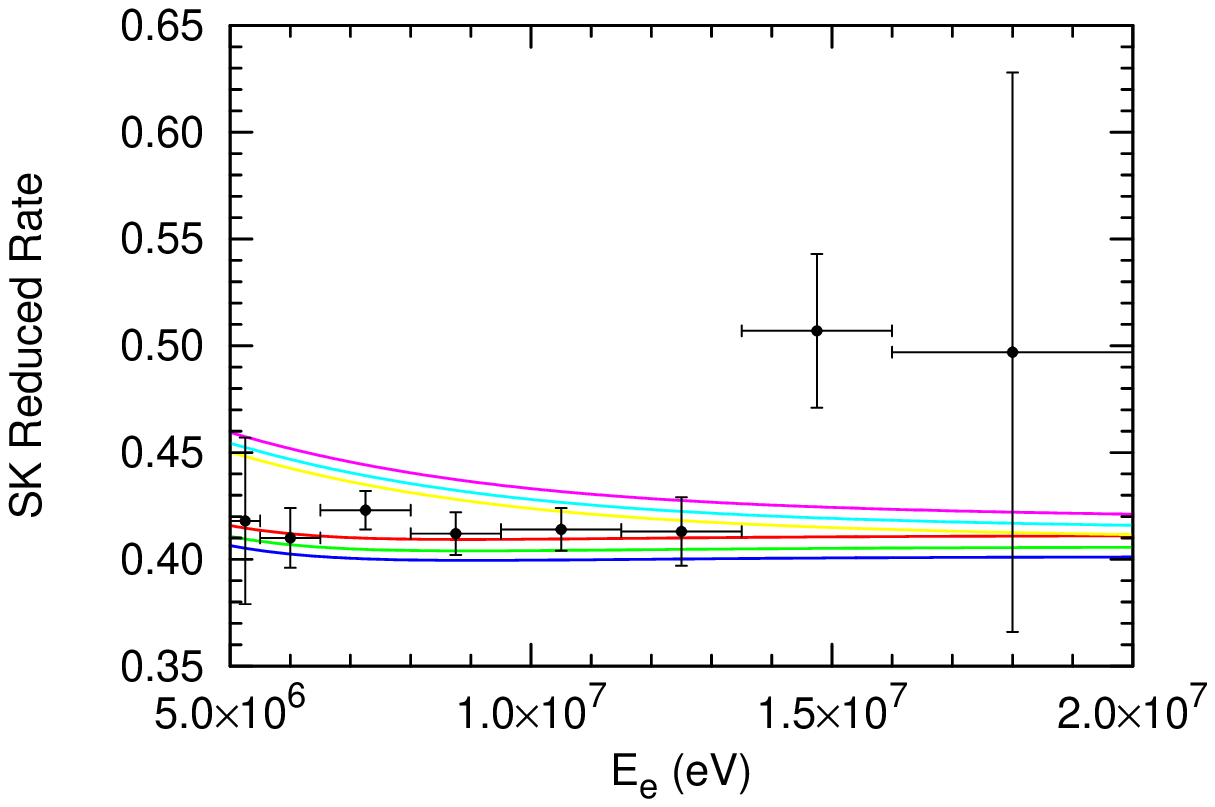}
\caption{\label{label}SuperKamiokande spectrum: data points and predictions
for vanishing field (upper curves) and $B_0=140kG$ (lower curves). From top to 
bottom and for each set $sin\theta_{13}=0,0.1,0.13$.}
\end{minipage}\hspace{2pc}%
\begin{minipage}{18pc}
\includegraphics[width=18pc]{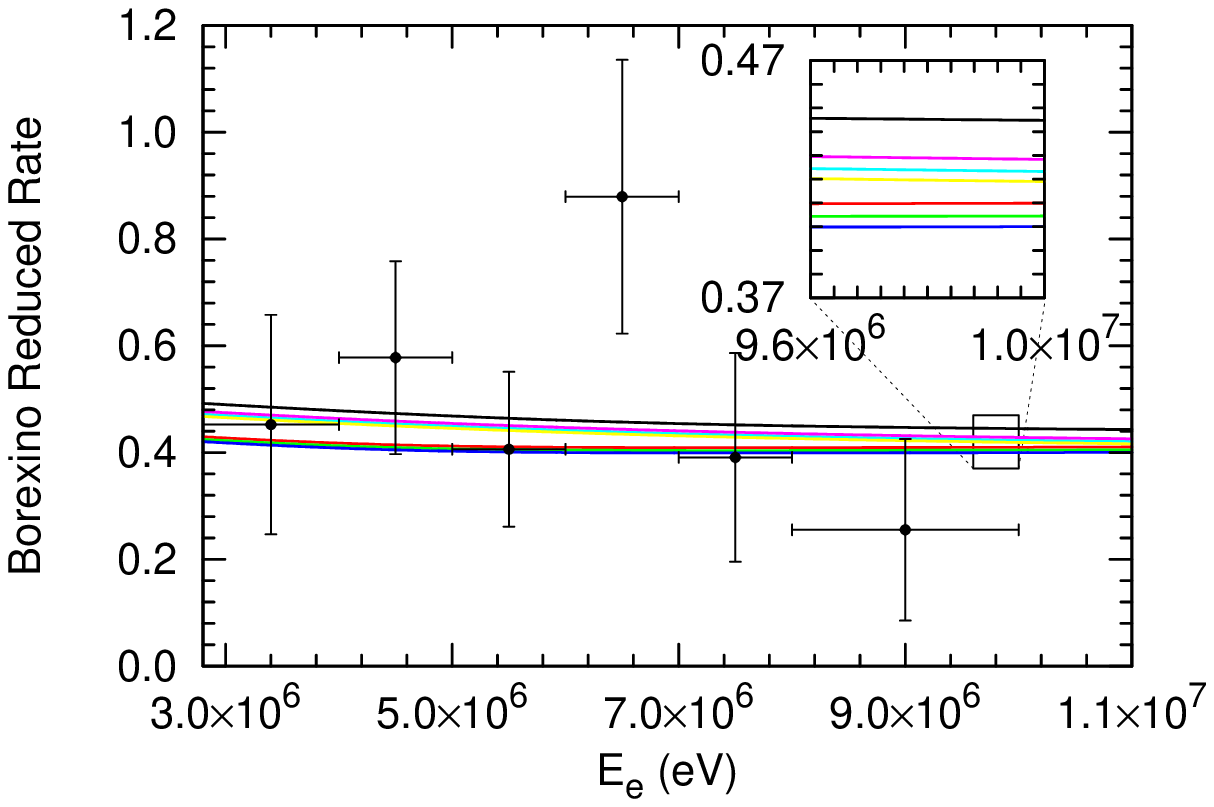}
\caption{\label{label}The same as fig.1 for Borexino spectrum and $^8 B$
neutrinos: all experimental errors are larger here and no conclusion can be 
drawn as for the preference of the data.}
\end{minipage}
\end{figure}
Since the intensity of the solar magnetic field is possibly related to the solar
activity, the time dependence of the data is an open possibility consistent with
this model. So we examine the periods in which the data were taken. We see that

\begin{itemize}
\item the SuperKamiokande spectrum refers to the period from May 31, 1996 to 
July 15, 2001 during which the average sunspot number was 65
\item the Borexino $^8 B$ spectrum refers to the period July 15, 2007 to 
June 21, 2008 when the average sunspot number was 4
\item in most of the former period the solar magnetic activity increased and
reached an 11 year peak in the Summer of 2000, whereas in the latter the
activity was continuously at its minimum.
\end{itemize}
Therefore in the light of this model one expects the Borexino $^8 B$
spectrum to coincide with the LMA prediction and the SuperKamiokande one
to reflect a moderately active sun. If and when Borexino are able to reduce 
their errors to 1/3 (2/3 reduction), solar activity will probably have increased.
Depending on whether this will be reflected in the increase of the solar
field strength and a consequent decrease in the event rate, an oportunity may 
arise to clearly test the model. 

Signatures of profiles 1 and 2 may also be distinguishable. In fact 
for profile 1 and Borexino phase 1 (low energy fluxes, $E<1.7MeV$) all $\nu's$  
have their resonances at $x<0.5$. Here the field is weak and the  
density is large so that the event rate modulation will be too small $(\simeq 1\%)$ 
to ever be seen. For profile 2 and Borexino phase 1, resonances lie 
at $x<0.23$ where the field is close to maximal $(O(1MG))$, so that a field modulation 
in connection with solar activity accounts for as much as $(\simeq 9\%)$ variability 
in the Borexino event rate. Hence low energy fluxes can detect modulation for profile 2 
but not for profile 1. Finally the $^8 B$ flux (seen by SK and Borexino) can detect 
modulation for both classes of profiles, so it cannot tell profile 1 from 2.

Hence we believe it essential to keep Borexino
taking data for all neutrino fluxes during at least the first half
of the present solar cycle expected to peak in 2011 or 2012
and present their data in time bins.

\smallskip

\end{document}